\documentclass[12pt]{elsart}
\usepackage{amssymb}
\usepackage{graphicx}
\begin{document}
\begin{frontmatter}
\title{
Projected multicluster model with Jastrow and linear state dependent 
correlations for $12 \leq A \leq 16$ nuclei.
}
\author[address1]{E. Buend\'{\i}a}
\ead{buendia@ugr.es},
\author[address1]{F. J. G\'alvez}
\ead{galvez@ugr.es},
\author[address2]{A. Sarsa\corauthref{cor1}}
\ead{fa1sarua@uco.es} 
\corauth[cor1]{Corresponding author}
\address[address1]{
Departamento de F\'{\i}sica Moderna, Facultad de Ciencias\\
Universidad de Granada, E-18071 Granada, Spain}
\address[address2]{
Departamento de F\'{\i}sica, Campus de Rabanales\\
Universidad de C\'ordoba, E-14071 C\'ordoba, Spain}

\begin{abstract}

Variational wave functions based on a Margenau-Brink cluster model
with short range and state dependent correlations, and angular
momentum projection are obtained for some nuclei with $12 \leq A \leq
16$.  The calculations have been carried out starting from the
nucleon-nucleon interaction by using the Variational Monte Carlo
method.  The configuration used consists of three alpha clusters
located at the apexes of an equilateral triangle, and an additional
cluster, not necessarily of alpha type, forming a tetrahedron.
This cluster is located at the top of its height.
Short-range and state dependent correlations are included by means of
a central Jastrow factor and a linear operatorial correlation factor
respectively.  Angular momentum projection is performed by using the
Peierls-Yoccoz operators.  Optimal structures are obtained for all the
nuclei studied.  Some aspects of our methodology have been tested by
comparing with previous calculations carried out without short range
correlations.  The binding energy, the root mean square radius, and
the one- and two-body densities are reported.  The effects of
correlations on both the energy and the nucleon distribution are
analyzed systematically.

\end{abstract}


\end{frontmatter}

\newpage

\section{Introduction}
\label{introduction}

The joint use of short-range dynamic correlations with model wave
functions including relevant aspects of the nuclear structure
constitutes the most commonly used scheme to describe nuclear bound
states with realistic or semi realistic interactions.  Short range
correlations are essential elements in the wave function because, as
it is well known, any of the so-called realistic or semi-realistic
parameterizations of the nuclear potential presents a strong
short-range repulsive core.  On the other hand, the formation of
different kind of clusters in the nuclei can be understood as a
collective movement of the nucleons governed by the medium and long
range part of the nuclear potential.  Therefore, for an accurate
description of the nuclear states, it is convenient to consider both
aspects in any variational approach to the nuclear bound states using
this type of interactions.  In principle, short range correlations are
mainly governed by the nucleon-nucleon interaction while medium and
long range effects depend on the particular nuclear state.  However,
and in a more careful approach, the final form of the short range
correlations will depend on the model wave function giving rise to a
non negligible dependence of the correlations on the nucleus.  

A direct way to include both short range and medium and long range
correlations is by using Jastrow type correlation factors, but
the calculation of the expectation values becomes very cumbersome,
especially when state dependent correlations are included. There exist
several methods to evaluate these expectation values as those based on
cluster expansions \cite{bishop78,bosca88}, the
Fermi-HiperNetted-Chain method \cite{fabro98,fabro00} or statistical
methods such as the Variational Monte Carlo
\cite{carlson88,pieper90}. The Coupled Cluster method allows to
incorporate both type of correlations
\cite{zabo81,feldmeier98,heisen99}. In this way it is possible to
understand how the different correlation mechanisms are incorporated
\cite{bishop90,guar96}

Alpha cluster models, or cluster models in general, have been widely
applied in microscopic descriptions of bound and scattering states of
nuclear systems \cite{dedu97,dude97}.  Variational wave functions
built within this framework constitute an appropriate scheme for
nuclei such as $^8$Be and $^{12}$C, that present a clear cluster
structure.  The use of wave functions including the possibility of
formation of alpha cluster structures or any other kind of grouping of
nucleons improves the description of these nuclei and their neighbours
with respect to simple mean field approximations
\cite{arima72,michel98}.

Multi cluster models have been used in microscopic calculations,
i.e. without effective cluster-cluster interactions, based on the
Generator Coordinate Method for some nuclei between A=12 and A=16
\cite{dude96,desc02}.  In these works a Volkov nucleon-nucleon
potential was used \cite{volk65}.  Other results of microscopic
multicluster calculations based on the stochastic variational method
have been reported \cite{vst95,oasv00} for some nuclei using the
Minnesota potential.  Neither of these potentials presents a strongly
repulsive short range part and, therefore, short range correlations do
not play a significant role.  On the other hand, previous studies of
alpha clustering based on nuclear potentials with a strongly repulsive
core have been mainly restricted to spin-isospin saturated nuclei
\cite{gmn01,bgps02}.

The aim of this work is to study the ground state of some $p$-shell
nuclei, $A\neq 4n$, including clustering effects and short range and
state dependent correlations, starting from $v_4$--type
nucleon-nucleon interactions.  The nucleon clustering is described in
terms of model wave functions based on a generalized Margenau-Brink
model as in \cite{dude96}.  Short range correlations are included by
means of a Jastrow factor and the dependence on the spin and isospin
exchange channels is included by using a linear state dependent
correlation factor.  Angular momentum projection is carried out in
order to obtain variational wave functions that are eigenfunctions of
the total angular momentum operator.  The calculations are performed
by means of the Variational Monte Carlo method.

Here we extend a previous work \cite{bgps02} to the $A\neq 4n$ case.
This generalization is not straightforward because the angular
momentum projection involves a spin mixing not present in spin and
isospin saturated nuclei.  In this paper we present an analytical
reduction of the different expectation values for these nuclei,
obtaining expressions suitable for the Variational Monte Carlo method.
By using this scheme, the computing time is hardly increased with
respect to the spin-isospin saturated case.  We apply the method to
the ground state of $^{13}$C, $^{14}$C, $^{14}$N, and $^{15}$N. 
The results obtained are also valid for the mirror nuclei
$^{13}$N, $^{14}$O and $^{15}$O because the electrostatic energy has
been not considered in the minimization process.  A systematic
analysis of the effects of the different correlations mechanisms
included in the wave functions on the total energy and on the
contribution of the different channels is carried out.  One and two
body densities are reported and the effect of the correlations are
discussed.

The scheme of this work is as follows. In Section 2 the variational wave
function and the analytical reduction of the expectation values leading to 
a form appropriate for the Variational Monte Carlo method are detailed.
In Section 3 we report and discuss the main results here obtained. 
The conclusions of the present work can be found in Section 4.

\section{Wave function}
\label{wavefunction}

The variational trial wave function used in this work is 
\begin{equation}
\Psi^{\pm}_{JKM}(1,2,\ldots, A)=
F_{\mathcal J}(1,\ldots,A)
F_{\mathcal L}(1,\ldots,A) 
\Phi^{\pm}_{JKM}(1,\ldots,A)
\end{equation}

This structure has been used in previous studies of spin and isospin
saturated nuclei \cite{bgps02,bgps01}.  It consists of a central
Jastrow correlation factor $F_{\mathcal J}$, a linear correlation
factor $F_{\mathcal L}$ that can include state dependent correlations,
and a model wave function $\Phi^{\pm}_{JKM}$ that is antisymmetric and has
the proper values of the total angular momentum and parity. 

The Jastrow factor depends only on the distance between pair of
nucleons
\begin{equation}
F_{\mathcal J}(1,\ldots,A)=\prod_{i<j}^A f(r_{ij}). 
\end{equation}
The linear factor is defined as
\begin{equation}
F_{\mathcal L}(1,\ldots,A)=\sum_{i<j}^A g(i,j)  
\end{equation}
where the function $g(i,j)$ depends on the radial and intrinsic 
degrees of freedom of particles, $i,j$. This is the only part of the
trial wave function where state dependent correlations are present
explicitly.  Here we employ the same parameterization for the
correlation functions $g(i,j)$ and $f(r)$, used in previous works
\cite{gmn01,bgps02,bgps01} that has shown to provide good
results
\begin{equation}
g(i,j)=\sum_{k=1}^{4}g^{(k)}(r_{ij}){\bf P}^{(k)}(i,j),
\end{equation}
where
\[
{\bf P}^{(1)}(i,j)=1,~~~
{\bf P}^{(2)}(i,j)=\frac{1}{2}\left(1+
\vec{\sigma}_i\cdot\vec{\sigma}_j\right)
\]
\begin{equation}
{\bf P}^{(3)}(i,j)=\frac{1}{2}\left(1+
  \vec{\tau}_i\cdot\vec{\tau}_j\right),~~~ {\bf P}^{(4)}(i,j)={\bf
  P}^{(2)}(i,j){\bf P}^{(3)}(i,j). 
\end{equation}

This operatorial dependence of the correlation factor is the same
as that of the nucleon-nucleon interactions considered in this work.
The functions $g^{(k)}(r)$, $k=1,..,4$, and $f(r)$ are parameterized
as a linear combination of Gaussians
\begin{equation}
g^{(k)}(r)=\sum_{m=1}^{M} a_m^{(k)}~e^{-b_m r^2},~~~
f(r)=1+\sum_{n=1}^{N} c_n~e^{-d_n r^2}.  
\end{equation}

The new aspects of treating $A\neq 4n$ nuclei with respect to spin and
isospin saturated ones are originated in the angular momentum
projection.  Therefore we shall focus here on the model part of the
wave function and on the angular momentum projection.  The correlation
factors are treated as in the spin and isospin saturated case.

\begin{figure}[ht]
\begin{center}
\includegraphics[scale=0.90]{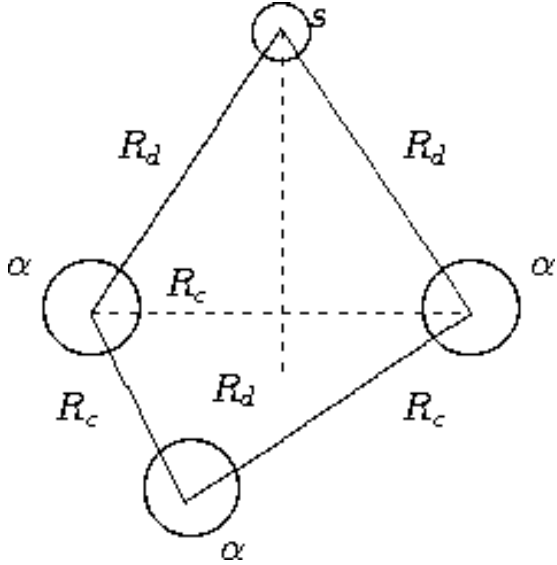}
\caption{\label{fig1}
Cluster description of the nuclei in terms of three alpha particles and
a general {\em s} incomplete cluster with 1,2 or 3 nucleons}
\end{center}
\end{figure}

The model wave function used here is based on a generalization
of the Margenau-Brink model.  Instead of using only alpha-particle
like nucleon clusters, more general groupings are allowed giving rise
to a  multicluster description \cite{dude96,vst95}. 
Within the molecular viewpoint of the Margenau-Brink scheme,
the model wave function is obtained starting from the following functions
\begin{equation}
\Phi_{\vec{{\bf C}}}(1,2,\ldots, A)=
{\mathcal  A} \left\{ 
\Phi_1(x_1,\ldots,x_{k_1})
\ldots \Phi_n(x_{k_{n-1}-1},\ldots,x_A) 
\right\}
\label{clusgen}
\end{equation}
where $\vec{\bf C}\equiv\left\{\vec{c}_k\right\}_{k=1}^n$ is a set of
parameters that represent the centers of the clusters, and ${\mathcal
A}$ is the corresponding antisymmetrizer.  In this work the
arrangement of the nucleons, shown in Fig. \ref{fig1}, consists of three
$\alpha$ clusters and a fourth incomplete cluster that can be made of
one, two or three nucleons depending on the nucleus under study.  

For this configuration, the general form of the function 
given in Eq. (\ref{clusgen}) reduces to 
\begin{equation}
\Phi_{\vec{{\bf C}}}(1,2,\ldots, A)=
{\mathcal  A} \left\{ \left[\prod_{m=1}^3
\Phi_{\alpha_m}(x_{4m-3},\ldots,x_{4m}) \right]
\Phi_s(x_{13},\ldots,x_A)
\right\}
\label{cluspart}
\end{equation}
where  $\Phi_{\alpha_m}$ stands for the wave function of an
alpha particle centered at $\vec{c}_m$, and $\Phi_s$ represents the
incomplete cluster wave function centered at $\vec{c}_s$.  

In this work the  $\Phi_{\alpha_m}$ functions are
taken to be Slater determinants built from harmonic oscillator single
particle orbitals centered at $\vec{c}_m$
\begin{equation}
\phi_{\beta, \, \vec{c}}(\vec{r})=
\left(\frac{\beta^2}{\pi}\right)^{3/4} \, 
e^{-\frac{1}{2} \beta^2 (\vec{r}-\vec{c})^2}
\end{equation}

The oscillator parameter, $\beta$, is the same for all of
the alpha clusters.  For the incomplete cluster wave function another
Slater determinant centered at $\vec{c}_s$ is employed also built from
$s$-wave harmonic oscillator single particle orbitals.  The oscillator
parameter in this case is, in general, different to that for the
$\alpha$ cluster wave function.  The importance of using a different
harmonic oscillator parameter will be discussed.  With these choices
for the cluster wave functions, the model wave function of the $A$
nucleons is a Slater determinant.  In general this function is not
eigenfunction of parity or total angular momentum operators.

The following linear combinations 
\begin{equation}
\label{paridad}
\Phi_{\vec{{\bf C}}}^{\pm}(1,2,\ldots, A)= 
\Phi_{\vec{{\bf C}}}(1,2,\ldots, A) \pm
\Phi_{-\vec{{\bf C}}}(1,2,\ldots, A) 
\end{equation}
have definite parity.  Model wave functions with the total angular
momentum of the state under study can be obtained from
Eq. (\ref{paridad}) by using the Peierls-Yoccoz projection operators
\cite{peyo57}
\begin {eqnarray}
\Phi^{\pm}_{JKM}(1,\ldots, A)
& = & \frac{2J+1}{8\pi^2}
\int d\Theta {\mathcal D}_{MK}^{J*}(\Theta){\bf R}(\Theta) 
\Phi^{\pm}_{\vec{\bf C}}(1,\ldots, A)
\end{eqnarray}
where ${\bf R}(\Theta)$ is the rotation operator, ${\mathcal
D}_{MK}^{J*}(\Theta)$ is the rotation matrix and $\Theta$ represents the
Euler angles. The quantum number $J$ gives the total angular momentum,
$K $ is its projection along the nuclear $z$ axis and $M $ the
projection along the $Z$ axis of the laboratory fixed frame. The
projection within this scheme is carried out by rotating the intrinsic
state and integrating over all angles weighted by the rotation
matrix. 

The function $\Phi_{\vec{\bf C}}(1,\ldots, A)$ in Eq. (\ref{cluspart})
is the generator function of the model wave functions.  Note that we
have removed the parametric dependence of the model wave function on
the position of the centers, $\vec{\bf C}$, in order to simplify the
notation. The distances between the clusters, $R_c$ and $R_d$, are
determined variationally.

The action of the rotation operator on the generator function is now
described in detail.  As we have mentioned before, this is the source
of the new methodological aspects originated by the fact that the
nuclear states are not spin and isospin saturated.  We do not need to
consider here the correlation factors because they are rotationally
invariant.  The generator function is a  Slater determinant.  The
action of the rotation operator on it leads to a
linear combination of Slater determinants.  If the Slater determinant
is spin and isospin saturated this linear combination contains only
one Slater determinant that also is spin and isospin saturated,
containing the same single-particle orbitals.  The only difference is
that, after rotation, these orbitals depend on the rotated
coordinates.  This was exploited previously to study $A = 4n$ nuclei
\cite{bgps01}.  When the nuclei are not spin or isospin saturated the
rotation gives rise to a mixing of spin states.

When the incomplete shell consists of one nucleon,
as for example in the ground state of $^{13}$C, 
the action of the rotation operator can be written as follows
\begin {eqnarray}
{\bf R}(\Theta)\Phi_{\vec{\bf C},\beta s_{\beta}t_{\beta}} & = & 
\sum_{s_i=\pm\frac{1}{2}}{\mathcal D}_{s_{\beta},s_i}^{\frac{1}{2}}(\Theta)
\overline{\Phi}_{\vec{\bf C},\beta s_i t_{\beta}}
\end{eqnarray}
where $\beta$ stands for the spatial quantum numbers of the orbital of the
incomplete cluster, and $s_\beta$ and $t_\beta$ are the third component 
of  spin and isospin, respectively.
The over line indicates that the Slater determinant must be evaluated on
the rotated coordinates.
Therefore, and concerning to the spin dependence of the state, the effect
of the rotation is to mix the two possible spin projections of the orbital
in the incomplete cluster. The weight of each component is given by the
matrix element of the rotation matrix.

When there are two extra nucleons the result of the rotation can be written
as follows
\begin {eqnarray}
{\bf R}(\Theta)
\Phi_{\vec{\bf C},\beta s_{\beta}t_{\beta},\beta s_{\gamma}t_{\gamma}} 
& = & 
\sum_{s_i,s_j=\pm\frac{1}{2}}
{\mathcal D}_{s_{\beta},s_i}^{\frac{1}{2}}(\Theta)
{\mathcal D}_{s_{\gamma},s_j}^{\frac{1}{2}}(\Theta)
\overline{\Phi}_{\vec{\bf C},\beta s_i t_{\beta},\beta s_j t_{\gamma}}
\nonumber\\
 & = & \sum_{s_i,s_j=\pm\frac{1}{2}} \sum_{S=0,1} 
\langle\frac{1}{2}\frac{1}{2}s_{\beta}s_{\gamma}
| S,s_{\beta}+s_{\gamma}\rangle\\
& &
\langle\frac{1}{2}\frac{1}{2}s_i s_j| S,s_i+s_j\rangle
{\mathcal D}^{S}_{s_{\beta}+s_{\gamma},s_i+s_j}(\Theta)
\overline{\Phi}_{\vec{\bf C},\beta s_i t_{\beta},\beta s_j t_{\gamma}}
\nonumber
\end{eqnarray}
where ($\beta s_{\beta}t_{\beta}$) and ($\beta s_{\gamma}t_{\gamma}$) 
stand for the quantum numbers of the orbitals of the incomplete shell.
Note that we have considered the same spatial dependence for both 
single particle orbitals. 
Therefore, if one is dealing with two extra protons ($^{14}$O) or two
extra neutrons ($^{14}$C) with the two possible spin orientations the
term  $S=1$ vanishes.
Only in the case of one proton and one neutron outside closed shell 
($^{14}$N) both total spin components will contribute.

Finally, the case of three nucleons outside closed shell ($^{15}$N and
$^{15}$O ) is a conjugate configuration to that of one nucleon outside
closed shell and it is handled in the same way.

The values allowed for $J$ and $K$ are governed by the symmetry group
of the system. i.e. by the spatial positions of the centers of the
clusters.  For the nuclei here considered the group is C$_{3v}$.  The
spin of the extra cluster must be also considered in determining the
possible values of $K$. If $M_S$ is the total spin third component the
allowed $K$ values are given by the selection rule $|K-M_S|=3n$, with
$n$ a positive integer \cite{dude96}, and, for any $K$, $ J \geq K$
and the parity is $\pi = (-1)^{J \pm S}$.  The energy grows with $K$,
providing different rotational bands.  In this work we are concerned
only with the ground state, therefore we shall restrict ourselves to
$K=1$ for $^{14}$N and $K=0$ for all the rest.  For one and three
extra nucleons $M_S=1/2$ and the ground state is $(1/2)^+$, and for
two extra nucleons there are two possibilities; i) both nucleons are
protons or neutrons $M_S=0$ and the state is $0^+$, ii) one nucleon is
a proton and the other a neutron $M_S=0,1$, and the $1^+$ ground state
must be constructed with $M_S = 1$ and $K=1$.

In order to compute the expectation value of the Hamiltonian in the
projected wave function it is convenient to use the following expression
\cite{bgps01,brin66}
\begin{equation}
\langle \Psi^{\pm}_{JKM}|{\bf H}|\Psi^{\pm}_{JKM}\rangle
=\frac{2J+1}{8\pi^2}
\int {\mathcal D}^{J*}_{MK}(\Theta)
\langle \Phi^{\pm}_{\vec{\bf C}}|
F_{\mathcal L} F_{\mathcal J}\, {\bf H} 
F_{\mathcal J} F_{\mathcal L}\,{\bf R}(\Theta)
|\Phi^{\pm}_{\vec{\bf C}}\rangle
\end{equation}

Let us focus on the spin-isospin configuration of the nuclear state.
Note that, because of the rotational invariance property of the
Hamiltonian, only the ket is rotated remaining the bra on its original
configuration. This is important because it determines the
configurations that gives non zero contribution to the integral when
projected onto the bra.  The action of the rotation operator is to
produce a linear combination of configurations containing the original
one.  One needs to analyze all of them to determine if, after the
action of the spin--isospin operators of $F_{\mathcal L}$ and the
Hamiltonian, the original configuration is obtained.  As a result,
only the original configuration appearing after rotation contributes
with both central and state dependent correlation factors, except
except for incomplete clusters made of one proton and one neutron with
$S=0$, that we have not studied here, for which two of the
configuration appearing after rotation give non zero contribution.
Note that the weight factor must be included when doing the integral
in all of the cases.  The treatment of state dependent correlations in
terms of the intermediate states is not modified with respect to the
case of spin and isospin saturated nuclei \cite{bgps00,bgps03}.

\section{Results}
\label{results}

First we will test the new methodological aspects implemented in this work
by comparing with the results of Dufour and Descouvemont \cite{dude96}
obtained by using a different computational scheme.  
We will employ for the test both the same nucleon-nucleon interaction 
(the Volkov V7 potential), and the same wave function as in \cite{dude96}.
It is worth to point out that the correlation factor is not needed
because the interaction does not present a strongly repulsive core.

In Table \ref{table1} we show for the ground state and some excited
states of the nuclei studied in this work the binding energy and the
root mean square radius, $\langle r^2 \rangle^{1/2}$.  As can be seen
from the table, both set of results are in a very good agreement.  The
spin-orbit interaction is not included in our work and therefore one
can not compare directly the results for nuclei with an odd number of
nucleons.  For these nuclei we have compared with the average value of
the states $1/2^-$ and $3/2^-$ of \cite{dude96}.  This average gives a
value that it is very close to the Monte Carlo result of this work,
specially for $^{13}$C where the spin-orbit splitting is smaller than
in $^{15}$N.  From this test it can be concluded that, for $A\neq 4n$,
the angular momentum projection scheme of this work 
provide reliable results. 

\begin{table}
\caption{\label{table1} Binding energy and root mean square radius,
$\langle r^2\rangle^{1/2}$, for different nuclear states calculated in
this work (mc) as compared with the results of Dufour and Descouvemont
(dd) \cite{dude96}.  Both calculations have been performed by using
the Volkov V7 interaction \cite{volk65} and the same variational wave
function without correlations.  The inverse of the oscillator
parameter, $\beta^{-1}$, and the distances between the clusters, $R_c$ and 
$R_d$, are also included.  The energies are in MeV, and $\langle
r^2\rangle^{1/2}$, $\beta^{-1}$, and $R_c$ and $R_d$, in fm.  The
statistical error in the Monte Carlo calculation is indicated in
parentheses.  The Coulomb energy has been included in the total energy.
}
\begin{tabular}{lllllll}
\hline
$^{A}$X$(K,J^{\pi})$                  & $\beta^{-1}$ & $R_c, R_d$
& $E_{mc}$  & $E_{dd}$ & $\langle r^2\rangle^{1/2}_{mc}$ 
& $\langle r^2\rangle^{1/2}_{dd}$ \\
\hline
$^{12}$C$(0,0^+)$                     & 1.38         & 2.65 
&  86.49(4)  & 86.7    & 2.31(7)      & 2.31 \\
$^{12}$C$(3,3^-)$                     & 1.38         & 3.14 
&  76.41(4)  & 76.5    & 2.49(9)      & 2.49 \\
$^{13}$C$(\frac{1}{2},\frac{1}{2}^-)$ & 1.39         & 2.29,2.114 
&  88.99(7)  &  89.6   & 2.25(9)      & 2.25 \\
$^{14}$C$(0,0^+)$                     & 1.39         & 2.26,2.057 
&  102.26(6) &  102.5  & 2.26(7)      & 2.26 \\
$^{15}$N$(\frac{1}{2},\frac{1}{2}^-)$ & 1.35         & 1.84,1.887 
&  119.37(7) &  121.9  & 2.15(11)     & 2.15 \\
$^{16}$O$(0,0^+)$                     & 1.34         & 1.49,2.409 
&  147.83(5) &  148.0  & 2.18(3)      & 2.18 \\
$^{16}$O$(3,3^-)$                     & 1.37         & 2.24,1.958 
&  129.46(10)&  129.8  & 2.27(10)     & 2.26 \\
\hline
\end{tabular}
\end{table}

The ground state of these nuclei has been studied in this work by
using a semi realistic potential.
We have used the modified Afnan-Tang nuclear potential
MS3 \cite{afta68,guar81}.  This is a $v_4$ type interaction with a
strongly repulsive core.  It gives meaningless results when used with
non correlated trial wave functions.  Thus, in order to analyze the effects
of nuclear correlations with respect to the non-correlated case, it is
more convenient to use an interaction with a less repulsive short
range part as the Brink-Boeker BB1 force \cite{brbo67}.

The ground state energy and the root mean square radius $\langle r^2
\rangle^{1/2}$ for different nuclei calculated from a number of trial
wave functions by using the BB1 and the MS3 interactions are reported
in Tables \ref{table2} and \ref{table2b}, respectively.  The optimal
parameters of the trial wave functions are also shown.  The notation
is as follows: MB stands for a non correlated trial wave function, JL
includes central Jastrow and linear state independent correlations and JLO is a
wave function with central Jastrow and state dependent linear
correlations.  In the JLO approach we have used the same non-linear
parameters as in JL, i.e. the variational freedom is restricted only
to the linear parameters of the different operatorial channels.  This
scheme has shown to work properly for spin and isospin saturated
nuclei \cite{bgps02,bgps01} in such a way that the loss of energy due
to this partial optimization was very small.  This is convenient
because when state dependent correlations are included, two things
happens; first the calculation becomes slower, and second, the
statistical error increases.  Therefore it
is very convenient, from a computational point of view, that the
non-linear parameters can be well determined by means of a state
independent optimization. Note that the linear parameters are computed  by
solving a generalized eigenvalue problem and then only a long run is
required to fix them.  The expectation value of the Coulomb energy
$E_c$, not included in the total binding energy, is reported separately.
 For the results shown in this work we
have used $2^8 \times 10^5$ ($2^6 \times 10^5$) moves per-nucleon with
state independent (state dependent) correlated wave functions.

\begin{table}
\caption{\label{table2} Ground state energies calculated by using
different trial wave functions without correlations (MB), with state
independent correlations (JL) and with linear state dependent
correlations (JLO) for the BB1 Brink-Boeker potential.  Energies are
in MeV, $\langle r^2\rangle^{1/2}$ in fm, $\beta_1, \beta_2$ in
fm$^{-1}$ and $R_c, R_d$, in fm.  The statistical error is shown in
parentheses.  The Coulomb energy is not included in the total energy.
}
\begin{tabular}{lllllll}
\hline
$^{A}$X$(K,J^{\pi})$ 
& WF            & $\beta_1, \beta_2$ & $R_c, R_d$ 
& $E$           & $E_{c}$              & $\langle r^2\rangle^{1/2} $ \\
\hline
& MB            & 0.70                 & 3.4 
& $-$80.01(4)   & 7.197(1)             & 2.63(4) \\
$^{12}$C$(0,0^+)$
& JL            & 0.72                 & 3.5
& $-$112.36(4)  & 7.417(1)             & 2.53(7)\\
& JLO           & 0.72                 & 3.5       
& $-$117.68(11) & 7.397(1)             & 2.53(7)\\
\hline
& MB            & 0.68, 0.59           & 3.5, 3.0
& $-$78.29(6)       & 7.057(1)             & 2.71(9) \\
$^{13}$C$(\frac{1}{2},\frac{1}{2}^-)$ 
& JL            & 0.72, 0.54           & 3.4, 3.0 
& $-$112.65(7)  & 7.558(1)             & 2.53(8)\\
& JLO           & 0.72, 0.54           & 3.4, 3.0 
& $-$119.8(2) & 7.613(2)             & 2.52(15)\\
\hline
& MB            & 0.69, 0.56           & 3.2, 2.5 
& $-$86.36(5)   & 7.363(1)             & 2.64(6) \\
$^{14}$C$(0,0^+)$
& JL            & 0.74, 0.58           & 3.1, 2.8 
& $-$122.93(8) & 7.836(1)             & 2.47(5)\\
& JLO           & 0.74, 0.58           & 3.1, 2.8 
& $-$131.75(13) & 7.854(1)             & 2.46(8)\\
\hline
& MB            & 0.68, 0.57           & 3.2, 2.8 
& $-$85.09(6)   & 9.849(1)             & 2.65(8) \\
$^{14}$N$(1,1^+)$
& JL            & 0.71, 0.57           & 3.0, 2.5 
& $-$121.68(7) & 10.438(1)             & 2.47(7)\\
& JLO           & 0.71, 0.57           & 3.0, 2.5 
& $-$131.8(2) & 10.381(2)             & 2.48(10)\\
\hline
& MB            & 0.66, 0.56           & 3.0, 2.5 
& $-$97.69(10)    & 9.948(1)             & 2.65(9) \\
$^{15}$N$(\frac{1}{2},\frac{1}{2}^-)$ 
& JL            & 0.74, 0.63           & 2.7, 2.4
& $-$139.55(10) & 10.821(1)            & 2.39(9)\\
& JLO           & 0.74, 0.63           & 2.7, 2.4 
& $-$152.0(4) & 10.837(5)            & 2.38(18)\\
\hline
& MB            & 0.66                 & 2.9, 2.4 
& $-$118.70(5)  & 13.470(1)            & 2.60(3) \\
$^{16}$O$(0,0^+)_{C_{3v}}$
& JL            & 0.76                 & 2.8, 2.4 
& $-$166.92(6)  & 14.516(1)            & 2.36(3)\\
& JLO           & 0.76                 & 2.8, 2.4 
& $-$179.46(10) & 14.515(2)            & 2.35(5)\\
\hline
& MB            & 0.67                 & 2.8 
& $-$118.52(5)  & 13.456(1)            & 2.60(3) \\
$^{16}$O$(0,0^+)_t$
& JL            & 0.74                 & 2.6 
& $-$166.66(6)  & 14.446(2)            & 2.37(4)\\
& JLO           & 0.74                 & 2.6
& $-$180.61(8)  & 14.552(2) & 2.35(5)\\
\hline
\end{tabular}
\end{table}

\begin{table}
\caption{\label{table2b} Ground state energies calculated by using
different trial wave functions without correlations (MB), with state
independent correlations (JL) and with linear state dependent
correlations (JLO) for the modified Afnan-Tang MS3 potential.
Energies are in MeV, $\langle r^2\rangle^{1/2}$ in fm $\beta_1,
\beta_2$ in fm$^{-1}$ and $R_c, R_d$, in fm.  The statistical error is
shown in parentheses.  The Coulomb energy is not included in the total
energy.
}
\begin{tabular}{lllllll}
\hline
$^{A}$X$(K,J^{\pi})$ 
& WF            & $\beta_1, \beta_2$ & $R_c, R_d$ 
& $E$           & $E_{c}$              & $\langle r^2\rangle^{1/2} $ \\
\hline
$^{12}$C$(0,0^+)$
& JL            & 0.70                 & 3.5 
& $-$74.54(5)  & 7.571(1)             & 2.48(4)\\
& JLO           & 0.70                 & 3.5 
& $-$87.2(4)    & 7.440(2)             & 2.49(15)\\ 
\hline
$^{13}$C$(\frac{1}{2},\frac{1}{2}^-)$ 
& JL            & 0.70, 0.46           & 3.3, 3.1 
& $-$73.37(10)  & 7.833(1)             & 2.47(8)\\
& JLO           & 0.70, 0.46           & 3.3, 3.1 
& $-$88.6(6)    & 7.864(1)             & 2.44(13)\\
\hline
$^{14}$C$(0,0^+)$
& JL            & 0.69, 0.48           & 3.4, 3.0 
& $-$77.52(7)   & 7.840(1)             & 2.50(5)\\
& JLO           & 0.69, 0.48           & 3.4, 3.0 
& $-$94.6(3)    & 7.840(1)             & 2.44(10)\\
\hline
$^{14}$N$(1,1^+)$
& JL            & 0.69, 0.54           & 3.2, 2.8 
& $-$81.95(9)   & 10.699(1)             & 2.42(7)\\
& JLO           & 0.69, 0.54           & 3.3, 3.8 
& $-$99.3(4)    & 10.865(3)             & 2.37(10)\\
\hline
$^{15}$N$(\frac{1}{2},\frac{1}{2}^-)$ 
& JL            & 0.67, 0.54           & 3.2,2.8 
& $-$91.77(12)    & 10.701(1)            & 2.45(9) \\
& JLO           & 0.67, 0.54           & 3.2, 2.8 
& $-$112.6(6)     & 10.878(3)          & 2.39(15)\\
\hline
$^{16}$O$(0,0^+)$
& JL            & 0.71                 & 2.7 
& $-$114.46(7) & 14.827(1)            & 2.32(3)\\
& JLO           & 0.71                 & 2.7 
& $-$135.6(3) & 15.036(2)            & 2.27(7)\\
\hline
\end{tabular}
\end{table}

The wave functions used in this work includes two different
oscillator parameters, one for the complete clusters and another
for the incomplete one. 
This gives rise to an improvement in the energy of about 3 or 4 MeV
when the incomplete cluster is made of one or two nucleons.
The improvement is noticeably reduced if the incomplete cluster
contains three nucleons.
The smaller value for the oscillator parameter of the incomplete cluster is 
due to the fact that the nucleons are more localized in the alpha particle
cluster than in the incomplete cluster.
In general we have obtained oscillator parameters that vary between those 
of $^{12}$C and $^{16}$O.

With respect to the optimum parameters of the inter-cluster distances,
we have obtained that the distance between the centers of the complete
clusters is bigger than the distance between the incomplete cluster
and an alpha-particle cluster.  The total energy is not very sensitive
to variations of the inter-cluster distances in the neighbourhood of
the equilibrium values.  We have indicated such situation by giving
these distances with only one decimal digit.  Finally and, as it could
be expected, when moving from $A=12$ to $A=15$ the optimal values of
the variational parameters tend to those of $^{16}$O.  This is the
case for all of the interactions and wave functions analyzed in this
work. It is remarkable that the ground state energy of $^{16}$O
obtained with the $C_{3v}$ symmetry is practically the same as the
one obtained with a tetrahedral symmetry.

In general, the effect of the correlations is to reduce the average
size of the nucleus.  Therefore, the optimum values in the model wave
function will depend on the presence, or not, of the correlation factor.
The modification with respect to the non-correlated
wave function is roughly proportional in all of the parameters in such
a way that nucleon correlations give rise to an isotropic contraction
of the nucleus.

It is interesting to point out the importance of correlations in the
binding energy of $^{12}$C and $^{14}$C as compared with $^{13}$C and
$^{14}$N, respectively. With both interactions, $^{12}$C is more
bounded than $^{13}$C with central correlations, but state dependent
correlations reverse this situation, obtaining a difference of 1 and 2
MeV with the MS3 and BB1 interaction, respectively. The behaviour of
the nuclear binding energy of $^{14}$C and $^{14}$N is different with
both potentials. With the BB1 interaction, and without correlations,
$^{14}$C is slightly more bounded than $^{14}$N. The difference in
their binding energy decreases with the use of central correlations
and is zero with state dependent correlations. However, with the MS3
potential, $^{14}$N is 4.5 MeV more bounded than $^{14}$C with central
and state dependent correlations. The reason of this different
behaviour lies in the contribution of the Bartlett and Heisenberg
channels of the MS3 interaction, that are null in the BB1
potential. Finally it is also worth mentioning here that  we have
obtained a negligible effect of the state dependent correlations on
the Coulomb energy,  which depends basically on the
parameters of the model wave function.

The correlations increase the binding energy by a quantity which grows
with the number of nucleons $A$.  In order to get a deeper insight into
the coupling between correlations and the particular nucleus we report
in Table \ref{table3} the increment in energy per number of pairs of
nucleons.  For example the increase in the binding energy per nucleon
pair when state independent correlations are included with respect to
the uncorrelated model is given by 
\[
\Delta_{\rm JL-MB }=
 \frac{2}{A(A-1)} \left( E_{\rm JL} -E_{\rm MB} 
\right)
\]
where $E_{\rm JL}$ ($E_{\rm MB}$) is the energy in the JL (MB) model.
The quantity $\Delta_{\rm JLO-JL }$ is defined in a similar way.
As it can be seen, the increment per number of pairs is roughly constant
for all of the nuclei considered, specially $\Delta_{\rm{JLO-JL}}$,
that accounts for the effect of state dependent correlations.
The increment due to state dependent correlations in the MS3
potential is practically twice the increment in the BB1 case.

\begin{table}
\caption{\label{table3}
Increase in the binding energy per number of nucleon pairs due to the
inclusion of different correlation factors for the nuclei studied
in this work.
In parentheses is indicated the nuclear interaction.
The increment is in MeV per number of nucleon pairs.
The error is in the last figure.
}
\begin{tabular}{llll}
\hline
$^{A}$X$(K,J^{\pi})$ 
& $\Delta_{\rm{JL-MB}}$(BB1) &  $\Delta_{\rm{JLO-JL}}$(BB1) 
& $\Delta_{\rm{JLO-JL}}$(MS3) \\
\hline
$^{12}$C$(0,0^+)$                     & -0.49 & -0.08 & -0.19 \\
$^{13}$C$(\frac{1}{2},\frac{1}{2}^-)$ & -0.44 & -0.09 & -0.19 \\
$^{14}$C$(0,0^+)$                     & -0.40 & -0.10 & -0.19 \\
$^{14}$N$(1,1^+)$                     & -0.40 & -0.11 & -0.19 \\
$^{15}$N$(\frac{1}{2},\frac{1}{2}^-)$ & -0.40 & -0.12 & -0.19 \\
$^{16}$O$(0,0^+)(C_{3v})$             & -0.40 & -0.12 & -0.18 \\
\hline
\end{tabular}
\end{table}

A more detailed analysis of the effect of the state dependent
correlations on the energy can be done by looking at the contribution
of the kinetic energy and of the different channels of the potential
energy. In Fig. 2 we plot the differences between these quantities
calculated with the JL and JLO wave functions for both the BB1 and the
MS3 interactions. Both the kinetic energy and the energy of the Wigner
channel rise with state dependent correlations for both
potentials. This increase is more important for the kinetic energy
with the MS3 potential than with the BB1 one, whereas the opposite
holds for the energy of the Wigner energy. For the BB1 potential, the
Majorana channel is the responsible for the decrease in the
ground state energy when state dependent correlations are considered. For
the MS3 interaction, the effect on the Majorana channel is practically
canceled with that on the kinetic and Wigner energies, and the
Bartlett and Heisenberg channels make the nuclei more bounded.  The
contribution of these two channels is very close and is nearly
independent of the nucleus considered.

\begin{figure}[ht]
\begin{center}
\includegraphics[scale=1.25]{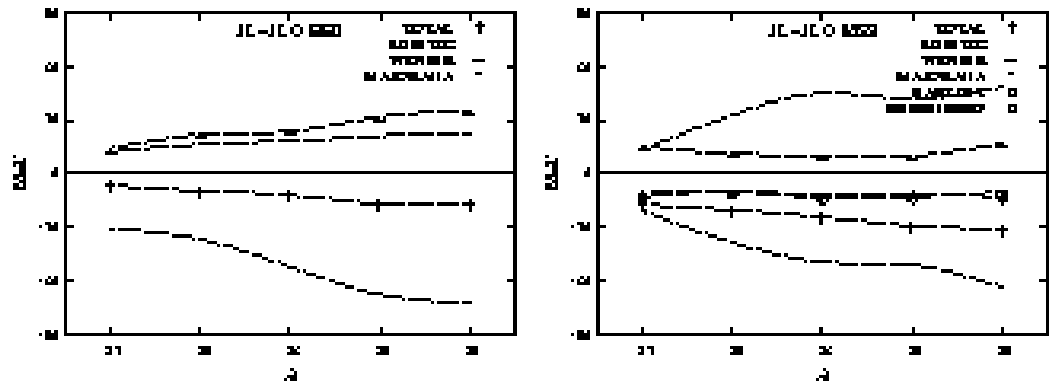}\hspace{0.1cm}
\caption{\label{fig2} Increase in the total energy, the expectation
values of the kinetic energy and the different channels of the
interacting potential when state dependent are included with respect
to the JL approximation.  In the left hand panel we plot the results
for the BB1 potential and in the right hand one for the MS3
potential. The lines are for guiding the eyes
}
\end{center}
\end{figure}

\begin{figure}[ht]
\begin{center}
\includegraphics[scale=1.25]{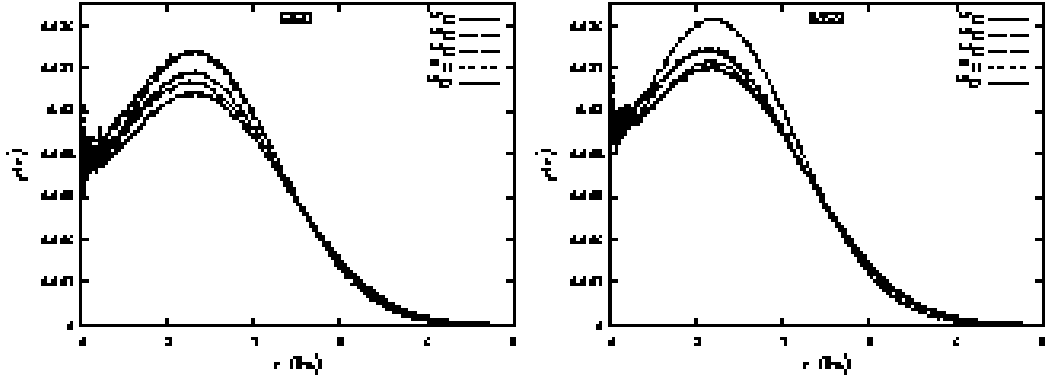}\hspace{0.1cm}
\caption{\label{fig3}
One body density for all the the nuclei studied in this work
calculated with the JL wave function.
In the left hand panel we plot the results for the BB1 potential and
in the right hand one for the MS3 potential.
}
\end{center}
\end{figure}

The one-- and two-- body densities give the spatial distribution of
the nucleons in the nuclei.
Here we have calculated these densities to analyze the effect of the
different correlation mechanisms introduced in the variational
wave functions.
In Fig.  \ref{fig3} we show the one body nuclear density 
calculated with the JL wave function for all of the
nuclei here studied and the two interactions considered.
As it can be seen, the qualitative behaviour is similar for both
potentials, with a higher value of the maximum as the number of
nucleons increases.
It is also worth pointing out that as $A$ increases, the density tends
to that of $^{16}$O.
It is for this nuclei and the MS3 interaction where this density is 
more separated from the others.

\begin{figure}[ht]
\begin{center}
\includegraphics[scale=1.25]{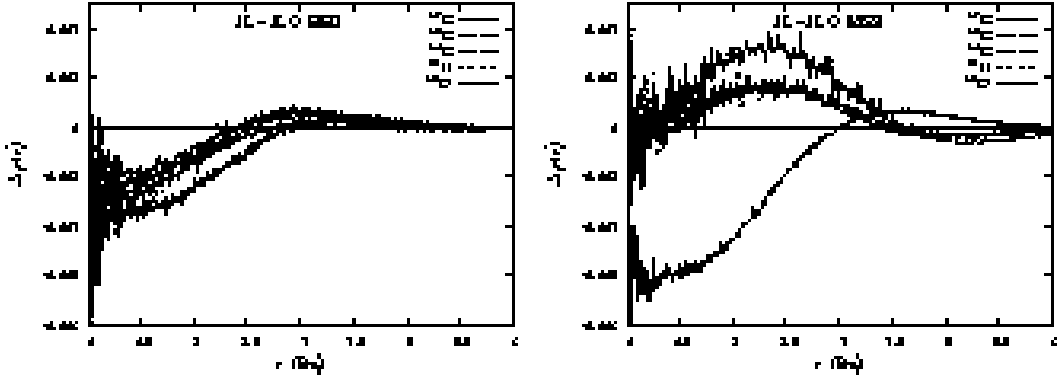}\hspace{0.1cm}
\caption{\label{fig4}
Effect of the state dependent correlations on the one body density
for the different nuclei considered in this work.
In the left hand panel we plot the results for the BB1 potential and
in the right hand one for the MS3 potential.
}
\end{center}
\end{figure}

The effect of the state dependent correlations on the one--body density
for these nuclei is studied in Fig. \ref{fig4}, where we plot the
difference between the single particle density obtained with the JL
and the JLO wave functions.  The first noticeable fact is that the
general behaviour is different for the two interactions used here.
Thus at short distances state dependent correlations tend to increase
the density with the BB1 interaction and the opposite happens with the MS3
potential, except for $^{12}$C for which a negative region at short
distances appears.  In addition, for the BB1 potential, the effect of
the operatorial correlations is roughly independent of the nucleus
while for the MS3 potential effects of the operatorial correlations
show a more accused dependence on the nucleus.

\begin{figure}[ht]
\begin{center}
\includegraphics[scale=1.25]{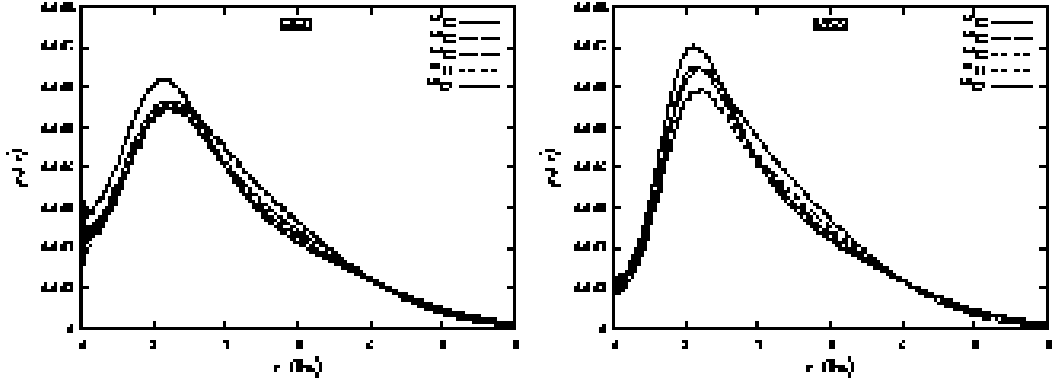}\hspace{0.1cm}
\caption{\label{fig5}
Two body density for all the the nuclei studied in this work
calculated with the JL wave function.
In the left hand panel we plot the results for the BB1 potential and
in the right hand one for the MS3 potential.
}
\end{center}
\end{figure}

The effects of correlations are more important on the two body density
than in the one body density.  In Fig. \ref{fig5} we plot the two body
density obtained from the state independent correlated wave function
JL for all of the nuclei studied and the two interactions considered
in this work.  The behaviour of this density is very similar for both
potentials, although the effect of the nuclear core is much more
important in the MS3 potential.  The main difference is that with the
MS3 interaction shorter distances are favoured with respect to the BB1
potential.  At distances between 2 and 3.5 fm the differences among
the nuclei considered are more important, with bigger values as the
number of nucleons increases from $^{12}$C to $^{16}$O.  This can be
understood as a progressive filling of the incomplete cluster that
gives rise to a larger number of particles at these intermediate
distances.

\begin{figure}[ht]
\begin{center}
\includegraphics[scale=1.25]{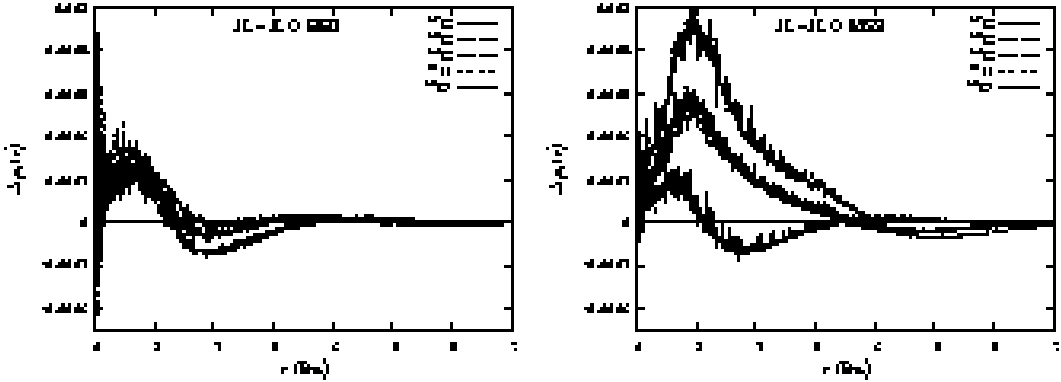}\hspace{0.1cm}
\caption{\label{fig6}
Effect of the state dependent correlations on the two body density
for the different nuclei considered in this work.
In the left hand panel we plot the results for the BB1 potential and
in the right hand one for the MS3 potential.
}
\end{center}
\end{figure}

Finally, the effect of including state dependent correlations on this
density is studied in Fig. \ref{fig6} where we plot the difference
between the two--body density calculated from the JL and JLO wave
functions.  As it was the case for the one body density, the effect of
state dependent correlations is roughly independent of the nucleus
when the BB1 potential is used and a more accused dependence is
observed for the MS3 interaction. For both potentials, state dependent
correlations bring together nucleons with respect to the JL case.

\section{Conclusions}
\label{conclusions}

Variational Monte Carlo calculations for $p$-shell, $A\neq 4n$, nuclei
starting from the nucleon-nucleon interaction have been presented.
The ground state energy and the one and two body densities have been
calculated.  The variational wave function consists of three factors:
a central Jastrow term, a spin-isospin dependent linear term and a model wave
function.  The model wave function is based on a cluster model
allowing for the formation of different kind of nucleon clusters with
centers at fixed positions.  The Peierls-Yoccoz projection operators
have been used in order to obtain trial wave functions with the proper
values of the angular momentum.  This work extend previous ones
carried out for spin and isospin saturated nuclei.

The present scheme has shown to be appropriate for describing two
important and complementary aspects of the nuclear dynamics as the
short range correlations and the formation of nucleon clusters.  The
former is induced by the short range repulsive part of the nuclear
potential while the later is a collective effect due to the medium and
long range part of the interaction.

In this work, an analytical reduction of the expectation values for
$A\neq 4n$ nuclei is presented. 
The use of the Peierls-Yoccoz projection operators introduces new 
features when the nuclei are not spin and isospin saturated.
Here we obtain a final form of the expectation values
which is specially suited for the Variational Monte Carlo calculation.
This is done for both state independent and state dependent correlation
factors.
As a result the different expectation values can be computed with
no significant extra computational cost with respect to the case of spin
and isospin saturated nuclei.

The scheme is applied to several nuclei with $12 \leq A \leq 16$.
The methodology has been first tested against previous works using
a completely different scheme of calculation.
Then results obtained by using two different nucleon-nucleon potentials 
including a repulsive core at short distances and state-dependent 
interaction channels have been reported.
The binding energies and the root mean square radius along with the
optimal parameters of the wave functions 
are shown for the different nuclei and states considered here. 
The effect of the different correlation mechanisms included in the trial
wave function on the energy and on the equilibrium geometries is
discussed.
The importance of using different oscillator parameters for the different
kind of nucleon clusters is shown.
The effect of the correlations on the different interaction channels
is analyzed in terms of the number of nucleons.
Finally one and two body densities obtained for the nuclei here studied
with several approximations of the wave functions are reported and
discussed.

\section*{Acknowledgements}

This work has been partially supported by the Ministerio de Ciencia y
Tecnolog\'{\i}a and FEDER under contract BFM2002-00200, and by the
Junta de Andaluc\'{\i}a.

\end{document}